\documentclass[twoside]{article}
\usepackage{fleqn,espcrc2}

\usepackage{graphicx}
\usepackage[figuresright]{rotating}

\newcommand{\Df}{D^{^{F}}}
\newcommand{\Do}{D^0}
\newcommand{\vfx}{v_{Fx}}

\newcommand{\be}{\begin{equation}}

\newcommand{\ee}{\end{equation}}
\newcommand{\bea}{\begin{eqnarray}}
\newcommand{\eea}{\end{eqnarray}}

\newcommand{\la}{\langle}
\newcommand{\ra}{\rangle}

\newcommand{\vr}{{\bf{r}}}
\newcommand{\vq}{{\bf{q}}}

\newcommand{\hj}{\hat{\alpha}}

\newcommand{\pll}{\parallel}

\newcommand{\AmS}{{\protect\the\textfont2
  A\kern-.1667em\lower.5ex\hbox{M}\kern-.125emS}}

\title{Quasiparticles and Phase Fluctuations in High Tc Superconductors}

\author{{Arun Paramekanti and Mohit Randeria}
 \address{
                Dept. of Theoretical Physics, Tata Institute of Fundamental Research, Mumbai 400005,
		India
	}}
       
\begin{document}

\begin{abstract}
We argue based on theoretical considerations and analysis of
experimental data that quasiparticle excitations near the nodes
determine the low temperature properties in the superconducting
state of cuprates. Quantum effects of phase fluctuations are shown
to be quantitatively important, but thermal effects are small for
$T \ll T_c$. An anisotropic
superfluid Fermi liquid phenomenology is presented for 
the effect of quasiparticle interactions on the temperature and doping
dependence of the low $T$ penetration depth.
\vspace{1pc}
\end{abstract}

\maketitle

\bigskip

{\section {INTRODUCTION:}}

\medskip

The superconducting (SC) state  of the high Tc
cuprates differs from conventional SCs in several ways: a d-wave
gap with low energy quasiparticle excitations near the nodes,
a small phase stiffness and a short coherence length.
There is some controversy about the importance of quasiparticles
\cite{lee97} versus phase fluctuations \cite{emery95} in determining
the low temperature properties. In this paper, we discuss this problem
focusing mainly on the doping and temperature dependence of the in-plane
superfluid stiffness $D_{_\pll}$ which is related to the penetration depth
$\lambda_{_\pll}$ through $\lambda^{-2}_{_\pll}=4\pi e^2
D_{_\pll}/\hbar^2 c^2 d_c$ where $d_c$ is the mean interlayer spacing; 
we will set $\hbar=c=e=1$ below.

\smallskip

We first review experimental
evidence for quasiparticle excitations at optimal doping. 
Transport data in the SC state in YBCO \cite{ong95,bonn92} 
shows a scattering rate decreasing sharply below $T_c$ implying long
lived quasiparticle excitations for $T \ll T_c$.
Direct evidence from ARPES in Bi2212 
shows the presence of sharp quasiparticle peaks 
over the {\em entire} Fermi surface\cite{arpes99} for $T \ll T_c$. 
Thermal conductivity data \cite{taillefer} in YBCO and Bi2212 shows 
$\kappa \sim T$ at low $T$. The slope predicted by 
quasiparticle theory \cite{kappa} is in good agreement \cite{taillefer}
with this
$\kappa$ data on Bi2212, using ARPES \cite{mesot99}
estimates for the Fermi velocity 
$v_{F}$ and the gap slope 
$v_{\Delta} = (2\hbar k_{_F})^{-1}d\Delta/d\phi$ 
at the node.
Experimentally, it thus seems that quasiparticle excitations exist
and are important at low temperature.

It then seems natural to interpret the linear $T$ dependence 
of $\lambda_{_\pll}(T)$ \cite{hardy93} as arising from nodal 
quasiparticles (QP). Ignoring QP interactions the layer stiffness
$D_{_\pll}(T) = D_{_\pll}(0) - A_0 T$ with 
$A_0=(k_{_B}\ln 2 /\pi) v_{F} /v_{\Delta}$.
ARPES estimates in Bi2212 \cite{mesot99} for $v_{F}$ and $v_{\Delta}$
give $A_0 \sim 0.8 meV/K$, whereas experiments at optimality 
measure a slope $\sim 0.3 - 0.4 meV/K$. Thus there is at least a factor
of two discrepancy which needs to be understood.

Alternatively, it has been suggested that this linear $T$
behavior could arise entirely from thermal phase fluctuations
\cite{classicalxy1,classicalxy2} without invoking nodal quasiparticles.
However, there are two reasons to believe that thermal phase fluctuations
are unimportant at low $T$ in the cuprates. (1) An effective action calculation 
for charged $d-$wave SCs \cite{arun99}, summarized below, shows that 
thermal phase fluctuations become important only 
near $T_c$ for Bi2212 at optimality. 
On the other hand, quantum phase fluctuations are important at low $T$
and suppress both  $D_{_\pll}(0)$ and the slope. 
For Bi2212, the resulting renormalized $D_{_\pll}$ is about 
$30\%$ smaller while the renormalized slope is about $25 \%$ 
smaller than the bare values. 
(2) Further, with underdoping, $D_{_\pll}(0)$ decreases \cite{uemura89}
and the slope of $D_{_\pll}(T)$ also shows evidence of
decreasing in Bi2212 and La214, although some YBCO data is consistent 
with a doping-independent slope (see the compilation in 
refs.~\cite{mesot99,xiang99}).
Insofar as the data indicate a doping dependent slope for $D_{_\pll}(T)$,
they independently rule out classical thermal phase fluctuations as the 
explanation \cite{classicalxy1,classicalxy2} 
for the linear $T$ dependence, since 
the slope of $D_{_\pll}$ in such 
theories is insensitive to doping. 
Finally there is another proposal for understanding $D_{_\pll}(x;T)$ invoking
incoherent pair excitations \cite{chen98} which, in our opinion,
does not properly include the effect of Coulomb interactions.

We next summarize our phase action calculation, and then
describe how quasiparticle interactions could account for the difference
between the free QP value and the measured slope and its doping dependence. 

\bigskip

{\section{PHASE FLUCTUATIONS:}}

\medskip

We have recently derived, by appropriate coarse-graining, 
a quantum XY model describing phase fluctuations 
in charged, layered $d-$wave SCs, starting with a lattice 
model of fermions. See ref.~\cite{arun99} for details
of this derivation and some of the discussion in this Section.

The phase action for layered SCs with in-plane lattice spacing
$a=1$ takes the form
\bea
& &S[\theta]=\frac{1}{8 T}{\sum_{\vq,\omega_n}}'~\frac{\omega^2_n \xi_0^2}
{\tilde{V}_\vq} |\theta(\vq,\omega_n)|^2 \\ \nonumber
&+&\frac{1}{4}
\int_0^{1/T}d\tau\sum_{\vr,\hj}
\Df_{_\pll}
\left[1-\cos(\theta_{\vr,\tau}-\theta_{\vr+\hj,\tau})\right]
\label{quantumxy}
\eea
where $\xi_0$ is the in-plane coherence length,
and $\Df_{_\pll}$ refers to the layer stiffness {\it without}
phase fluctuation effects, but including possible renormalizations
due to quasiparticle interactions.
Here $\tilde{V}_\vq = V(\vq_{_\pll}/\xi_0,\vq_{_\perp})$ with
$V(\vq) = (2\pi e^2/q_\pll\epsilon_b)
\sinh(q_\pll d_c)/\left[{\rm cosh}(q_\pll d_c)-\cos(q_\perp
d_c)\right] $
is the Coulomb interaction for layered systems, $\epsilon_b$ is
the background dielectric constant, $d_c$ the interlayer spacing,
and $\vq_{_\pll},\vq_{_\perp}$ refer to in-plane and $c$-axis momentum 
components.
The prime on the sum indicates a Matsubara frequency cutoff 
since the energy of the fluctuations should not exceed the
condensation energy $E_{\rm cond}=\frac{1}{8}\Df_{_\pll}(\pi/\xi_0)^2$.
The form of the first term of (1), which arises from
coarse-graining up to a scale $\xi_0$, and the importance of cutoffs have 
not been appreciated earlier.
The (typically very small) $c$-axis stiffness $\Df_{_\perp}$ 
can be ignored for in-plane properties since it was found not to lead to 
qualitative or quantitative changes.

We ignore vortices (transverse phase fluctuations) which are suppressed at 
low $T$ by their finite core energy.
Analyzing longitudinal phase fluctuations for (1)
within a self-consistent harmonic approximation (SCHA)
\cite{classicalxy1}
leads to the renormalized 
stiffness $D_{_\pll} = \Df_{_\pll}\exp(-\la \delta\theta^2 \ra/2)$
where $\delta\theta^2=(\theta_{\vr,\tau}-\theta_{\vr+\alpha,\tau})^2$.
Our numerical results can be simply understood as follows: 
$\la \delta\theta^2 \ra (T=0) \sim
\sqrt{(e^2/\epsilon_b\xi_0)/D_{_\pll}(0)}$ is a measure of 
zero point quantum fluctuations, while
classical thermal phase fluctuations become important near a crossover
scale $T_\times \sim \min\left[T_c,T^0_\times\right]$ where
$T^0_\times = \sqrt{D_{_\pll}(0)(e^2/\epsilon_b\xi_0)}$
is the $T=0$ oscillator level spacing in the
renormalized harmonic theory.

It is easy to see that phase fluctuation effects are
negligible in the BCS limit, except very close to $T_c$.
With $e^2/\epsilon_b a \sim \Df_{_\pll} \sim E_{_F}$
and $\xi_0\sim v_{F}/\Delta$, one obtains the standard result 
$E_{\rm cond} \sim \Delta^2/E_{_F}$ per unit cell, and
$\la \delta\theta^2 \ra (T=0) \sim \sqrt{\Delta/E_{_F}} \ll 1$ and 
$T_\times \sim \min\left[T_c,\sqrt{E_{_F}\Delta}\right] = T_c$. 

For the cuprates, the small $\xi_0$ and small $\Df_{_\pll}$ act 
together to increase $\la \delta\theta^2\ra$, but they push $T_\times$ in 
opposite directions.
For optimal Bi2212 we use $e^2/\epsilon_b a \approx 0.33 eV$ 
with $\epsilon_b \approx 10$, $\xi_0/a \approx 10$, and $d_c/a \approx 4$.   
Assuming that the two layers within a bilayer are phase-locked, we get the 
bilayer 
stiffness $D_{_\pll}(0) \approx 80 meV$, from experimental data which
shows $\lambda_{_\pll} (0)\approx 2000 A$. This leads to $E_{\rm cond} 
\approx 6 
K/{\rm unit cell}$ and 
$T^0_\times \approx 600 K$. 
Since the bare stiffness $\Df_{_\pll}$ actually decreases
with temperature due to quasiparticle excitations,
an estimate of the crossover scale $T_\times$ can be
obtained from $T_\times \sim \sqrt{D_{_\pll}(T_\times)(e^2/\epsilon_b
\xi_0)}$. Assuming a linearly decreasing $D_{_\pll}(T)$, this leads to 
$T_\times \sim T_c$.
Thus longitudinal thermal fluctuations are clearly unimportant at 
low temperatures. 
Quantum fluctuations are important at low temperatures since 
$\la \delta\theta^2 \ra(T=0) \sim 1$ at optimality and detailed
calculations \cite{arun99} lead to 
a $30\%$ decrease of $\Df_{_\pll}(0)$ and a $25\%$ decrease in
the slope.

While it might appear that there could be a low $T$ crossover to 
thermal phase fluctuations due to the low energy $c$-axis plasmon 
($\sim 7 K$ for Bi2212) in the anisotropic layered SCs, the phase space for 
these low lying fluctuations is too small to lead to a linear $T$ behavior
\cite{arun99}. 
Even in a purely 2D system with a low lying $\sqrt{q_{_\pll}}$ 
plasmon, the decrease in the phase stiffness due to phase fluctuations 
only goes as a large power law ($\sim T^5$). Quasiparticles are 
thus crucial in obtaining the observed linear temperature dependence.

We next turn to the doping dependence of phase fluctuations.
The (amplitude) coherence length $\xi_0$ is crucial in determining 
the effect of phase fluctuations. Since $\xi_0$
is determined by the pairing gap which appears to remain finite
as we underdope, we do not expect singular behavior in the phase
fluctuations arising from the doping dependence of $\xi_0$.
In this case, the dominant doping dependence to phase fluctuations arises
only from the singular behavior of the bare parameters in the phase action
on underdoping. This singular $x$-dependence in $\Df_{_\pll}(x)$ is most 
naturally explained by quasiparticle interaction effects as discussed below.

To estimate the doping dependence of phase fluctuations, we note
that the core energy will lead to vortices being exponentially suppressed 
at low $T$, even as we underdope. Using the experimental input
$T_c \sim D_{_\pll}(x;T=0) \sim x$ \cite{uemura89} and
assuming a doping independent $\xi_0$, longitudinal fluctuations 
within the SCHA lead to $T^0_\times \sim \sqrt{x}$ and hence $T^0_\times 
\gg T_c $. Thus,
$T_\times \sim \min\left[T_c,T^0_\times\right] 
\sim T_c$, which implies that thermal
phase fluctuations are unimportant at low $T$ as one underdopes.
Further, within the SCHA, $\la \delta\theta^2 \ra \sim x^{-1/2}$, which
would lead to a destruction of superconductivity at small enough $x$. However,
we do not expect the SCHA to be valid close to this transition.

\bigskip

{\section {QUASIPARTICLE INTERACTIONS:}}

\medskip

The increasing importance of interactions with underdoping is
evident: $D_{_\pll}(x;0) \sim x$ \cite{uemura89}
and the quasiparticle weight diminishes \cite{jcc99} as one
approaches the Mott insulator. We thus explore the possibility that
residual interactions between the quasiparticles in the SC state
can account for the value and doping dependence of the slope
of $D_{_\pll}(x;T)$. To this end we use a phenomenological
superfluid Fermi liquid theory (SFLT) \cite{leggett75,millis98}.
All available experimental evidence on the ground state and low lying
excitations suggests that the correlated SC state in the cuprates is 
adiabatically connected to a d-wave BCS state. 
We thus feel that SFLT may be a reasonable 
description of QP interactions in the SC state at low $T$, even though this 
formulation makes reference to a (hypothetical) $T=0$ normal 
Fermi liquid in which SC is induced by turning on a pairing interaction. 
(One could argue that approaching the superconducting phase from 
the overdoped side at $T=0$, one obtains a normal Fermi liquid to SC
transition.)

The bilayer stiffness after including QP interaction effects is given
by, $\Df_{_\pll}(T)=\beta_{_F} \Do_{_\pll}(0) 
- \alpha_{_F} 2 (k_{_B} T \ln 2 /\pi) v_{F} /v_{\Delta}$ 
where $\alpha_{_F},\beta_{_F}$ are Fermi liquid renormalizations. 
We will constrain the Landau QP interaction function by demanding $\beta_{_F} 
\sim x$, consistent with experiments
and then determine the doping trends in $\alpha_{_F}$.

To compute $\alpha_{_F},\beta_{_F}$, 
using a standard Kubo formula in the quasiparticle basis,
it is convenient to shift the origin of the
Brillouin zone to the $(\pi,\pi)$ point and
describe the hole-like Fermi surface of Bi2212 in terms of an angle $\phi$.
The Landau $f$-function is denoted by $f(\phi,\phi')$.
We define $\la O \ra_{\phi} \equiv \int_0^{2\pi} 
d\phi k_{_F}(\phi) O(\phi)/[2\pi |v_{F}(\phi)|]$. 
We get
$\beta_{_F}= 1+ 4\pi \la\la \vfx (\phi) \vfx (\phi ')
f(\phi,\phi ') \ra\ra_{{\phi}{\phi\prime}}/\la \vfx^2 \ra_{\phi}$ 
from the diamagnetic response of the free energy 
$\partial^2 \delta F / \partial {\bf A}_x^2$ to an applied
vector potential.
The current carried by nodal quasiparticles is then renormalized by
the factor
$\sqrt{\alpha_{_F}}= 1 + \la \vfx(\phi) f(\phi_n,\phi)\ra_{\phi}
/\left[ \pi \vfx (\phi_n) \right ]$, 
relative to its non-interacting value, where the nodes are at
$\phi_n=(2n-1)\pi/4$ with $n=1\ldots 4$.

\begin{figure}
%\vskip -8mm
\vspace*{1.5ex}
\centerline{\includegraphics[angle=-90,width=3.0in, clip]{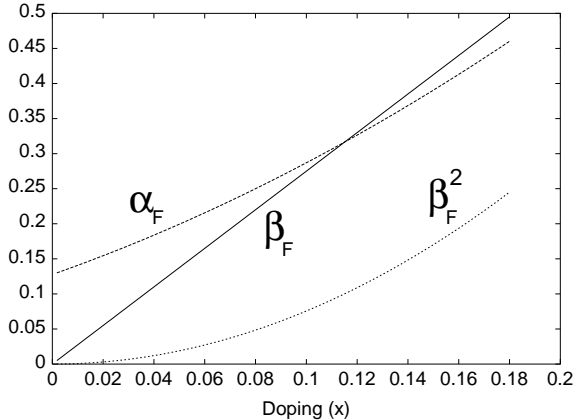}}
\vskip -1cm
\caption{\small Doping dependence of renormalizations 
$\alpha_{_F}$ and $\beta_{_F}$ plotted for anisotropic model discussed
in the text. For the isotropic case one gets
$\alpha_{_F}=\beta^2_{_F}$.}
\vskip -7mm
\end{figure}

We expand $f(\phi,\phi')= \sum_{m \geq m'} F_{m,m'} [ \cos(m \phi+m'\phi') 
+ \cos(m'\phi+m'\phi ') ]$ in a set of complete basis functions
\cite{aoi73},
where $m,m'=0,\pm 1, \pm 2, \ldots$ with square lattice symmetry
imposing $m+m'=4 p$ with $p=0,1,2,\ldots$.
In an isotropic system only $p=0$ survives {\it and}
$k_{_F}$ and $v_{_F}$ are $\phi$-independent. However, as
emphasized in ref.~\cite{millis98} one then obtains 
$\alpha_{_F} = \beta^2_{_F} \sim x^2$ in disagreement
with experiments \cite{mesot99}.

To illustrate how anisotropy can qualitatively change this scaling
we keep only the leading $p=0$ term: 
$f(\phi,\phi') = 2 F_{1,1} \cos(\phi-\phi ')$, 
but retain the full anisotropy of the dispersion
seen in ARPES \cite{arpes95}. 
We make a reasonable choice of $F_{1,1} = P + Qx$, with $P$ such that 
$\beta_{_F} \sim x$ as $x\to 0$,
and $Q$ such that $\beta_{_F} = 0.5$ at $x=0.2$. 
This leads to $\alpha_{_F}(x)$ shown in Fig.1, which
is a weak function of doping.
In general there are too many free parameters in the anisotropic case (an infinite 
set $F_{m,m'}$) for the theory to have predictive power; nevertheless the simple
example above shows how the $T=0$ value and slope of $D_{_\pll}$
can easily exhibit rather different $x$-dependences, and account for the
experimentally observed $D_{_\pll}(x;T)$.

We thus arrive at the following picture for the doping and temperature
dependence of $D_{_\pll}$. The bare stiffness arising from
non-interacting quasiparticles is renormalized by both QP interactions and 
quantum phase fluctuations at low $T$ leading to the measured
stiffness, $D_{_\pll}(x;T)$. Its doping dependence,
$D_{_\pll}(x;0) \sim x$, is determined by QP interactions while its linear $T$
behavior is governed by nodal QPs, with its slope
renormalized by both QP interactions and quantum phase fluctuations. 

\bigskip

{\bf Acknowledgments:} We thank J.C. Campuzano, J. Mesot, M.R. Norman,
C. Panagopoulos, T.V. Ramakrishnan and L. Taillefer for useful 
conversations. M.R. thanks the Indian DST for partial support through a 
Swarnajayanti fellowship.

\bigskip

%%%%%%%%%%%%%%%%%%%%%%%%%%%%%%%%%%%%%%%%%%%%%%%%%%%%%%%%%%%%%%%%%%%%%%%%%%%

\end{document}